\documentclass[prb, aps, twocolumn, groupedaddress, eqsecnum, superscriptaddress, nofootinbib, floatfix]{revtex4-2}

\usepackage[dvipsnames]{xcolor}
\usepackage{graphicx}
\usepackage[normalem]{ulem} 
\usepackage{enumitem}       

\usepackage{amsmath}
\usepackage{amssymb}
\usepackage{bbm}            
\usepackage{newtxtext,newtxmath} 
\usepackage{bm}             

\usepackage[version=4]{mhchem} 
\usepackage{hhline}
\usepackage{array}

\usepackage[colorlinks=true,linkcolor=Blue,citecolor=Blue,urlcolor=Blue]{hyperref}
\usepackage[capitalise]{cleveref}      

\begin{document}
\title{Fork Tensor Network Solver for Cluster Dynamical Mean-Field Theory}

\author{Tae-Young \surname{Lee}}
\affiliation{Department of Applied Physics, Graduate School, Korea University, Sejong 30019, Korea}

\author{Jeong-Hyeok \surname{Cha}}
\affiliation{Department of Physics and Astronomy, Seoul National University, Seoul 08826, Korea}

\author{Heung-Sik Kim}
\email{heungsikim@kentech.ac.kr}
\affiliation{Department of Energy Engineering, Korea Institute of Energy Technology, Naju-si 58330, Republic of Korea}

\author{Hyun-Yong \surname{Lee}}
\email{hyunyong@korea.ac.kr}
\affiliation{Division of Semiconductor Physics, Korea University, Sejong 30019, Korea}
\affiliation{Department of Applied Physics, Graduate School, Korea University, Sejong 30019, Korea}

\begin{abstract}
Real-frequency spectral functions of multiorbital clusters remain a challenge for the impurity solvers of cluster dynamical mean-field theory. We present a zero-temperature impurity solver for cluster dynamical mean-field theory based on fork tensor product states, in which the bath of each orbital and spin forms a separate arm attached to the cluster. 
The solver combines complex-time evolution with controlled bond expansion, which we extend from matrix product states to fork tensor product states. Controlled bond expansion avoids costly two-site updates of the cluster tensors and thereby speeds up the time evolution of a two-site cluster by almost an order of magnitude. Complex-time evolution slows the growth of entanglement, so that long times become accessible. Using hundreds of bath sites and a matrix-valued exponential fit of the complex-time data, with a projection that keeps the spectral function positive semidefinite, we close the self-consistency directly on the real-frequency axis without analytic continuation of imaginary-axis data.
For two- and three-orbital single-site dynamical mean-field theory, the spectral functions agree with recent complex-time results. For the two-orbital dimer of the two-plane Bethe lattice, we obtain the antiferromagnetic solution at zero temperature, whose main spectral features appear at positions similar to those of finite-temperature continuous-time quantum Monte Carlo results and are resolved as distinct peaks on the real-frequency axis. 
Our work thus provides a route to studying how short-range correlations affect high-energy features in the real-frequency spectra of correlated multiorbital materials.

\end{abstract}

\date{November 30, 2026}
\maketitle

\section{Introduction}

Dynamical mean-field theory (DMFT) is the standard framework for strongly correlated electron systems~\cite{Metzner1989,Georges1996,Kotliar2006}. It maps the lattice problem onto a quantum impurity model embedded in a bath that is determined self-consistently, and thereby captures local electronic correlations; in its multiorbital formulation it describes the physics of Hund's coupling that governs many transition-metal compounds~\cite{Georges2013}. Spatial correlations between neighboring sites, however, are beyond a single-site description. Cluster dynamical mean-field theory (CDMFT) includes the short-range spatial correlations within a cluster by embedding a cluster of sites rather than a single site in the bath~\cite{Maier2005,Kotliar2006}. The simplest cluster with orbital degrees of freedom, a dimer with two orbitals per site, is the minimal setting in which intersite correlations and Hund's coupling act together. The Hubbard model on the Bethe lattice composed of dimers, known as the two-plane Hubbard model, has mostly been studied with one orbital per site at half filling~\cite{Moeller1999,Fuhrmann2006,Kancharla2007,Hafermann2009}. With two orbitals per site, the dimer has been studied on the same lattice~\cite{Harland2019} and as a two-site cluster of the two-band Hubbard model~\cite{FlorezAblan2025}, and it describes dimerized transition-metal compounds in which molecular-orbital formation, orbital-selective behavior, and competing magnetic orders emerge~\cite{Harland2019}.

The central quantity of DMFT and CDMFT is the single-particle spectral function, and the computational bottleneck is the impurity solver that provides it. Continuous-time quantum Monte Carlo methods~\cite{Werner2006,Gull2011} treat multiorbital impurities at finite temperature, but they operate on the imaginary axis, so that real-frequency spectra require an ill-posed analytic continuation that broadens the high-energy features, and for multiorbital clusters they are increasingly affected by the fermionic sign problem. Exact diagonalization~\cite{Caffarel1994,Capone2007} works at zero temperature directly on the real-frequency axis, but it is limited to a small number of bath sites, so that the results depend on how the bath is parametrized and fitted~\cite{Senechal2010,FlorezAblan2025,Kim2026}. The numerical renormalization group~\cite{Wilson1975,Bulla2008} resolves the low-energy physics, but it is restricted to few orbitals and to a coarse resolution at high energies. Impurity solvers based on matrix product states~\cite{White1992,Schollwoeck2011} are free of the sign problem, work at zero temperature on the real-frequency axis, and are not restricted to a small number of bath sites, so that the bath spectral function can be represented almost exactly~\cite{Ganahl2015,Wolf2015,Bauernfeind2017,Cao2021}. Their applications to cluster problems, however, have remained scarce~\cite{Wolf2015,NunezFernandez2018}, and the matrix-valued Green's functions and hybridization functions of a cluster pose demands that single-site solvers do not face.

Two limitations of these tensor-network solvers become severe for multiorbital clusters. First, the spectral function follows from a real-time evolution whose reachable time, and hence frequency resolution, is bounded by the growth of entanglement with time, which forces the bond dimension to grow~\cite{Grundner2024,Yu2026}. Second, the geometry of the network determines the cost of the updates. In the fork tensor product states of Bauernfeind \textit{et al.}~\cite{Bauernfeind2017}, the bath of each orbital and spin is attached as a separate arm to the impurity tensor with which it hybridizes, so that the network reflects the connectivity of the impurity Hamiltonian; the same construction carries over to a cluster, where a bath couples only to the cluster sites of its own orbital and spin. The price to pay is that the impurity tensors carry several virtual legs. Their updates dominate the cost of both the ground-state search and the time evolution~\cite{Bauernfeind2017,Grundner2025}, and the two-site update that is normally required to let the bond dimension grow becomes prohibitively expensive at these tensors, because it scales with an additional power of the bond dimension~\cite{Grundner2025}. The remedy pursued so far has been to restructure the network itself by inserting branching nodes that carry only virtual legs~\cite{Gunst2018,Grundner2025}.

Two recent developments, both formulated for matrix product states, address precisely these two limitations. Controlled bond expansion~\cite{Gleis2023} enlarges the bond of a single-site update by those parts of the two-site space that carry significant weight, and thereby reaches the accuracy and convergence of two-site density matrix renormalization group updates at single-site cost; the same expansion turns the one-site time-dependent variational principle~\cite{Haegeman2011,Haegeman2016} into a rank-adaptive integrator~\cite{Li2024}. For a matrix product state the gain is a factor of order the local dimension; for a tensor with several virtual legs, where the two-site update costs an additional power of the bond dimension, it becomes decisive. Complex-time evolution~\cite{Cao2024,Grundner2024,Yu2026,Cha2025}, on the other hand, propagates the state along a contour in the complex-time plane, so that high-energy excitations are exponentially damped and the entanglement grows far more slowly than in real time. The real-frequency spectrum is then recovered from the complex-time data by an exponential fit in the pole representation of the Green's function~\cite{Zhang2024,Yu2026}.

In this work, we present a zero-temperature impurity solver for cluster dynamical mean-field theory that combines controlled bond expansion with complex-time evolution on fork tensor product states. The solver treats the matrix-valued Green's functions of multiorbital impurity clusters with hundreds of bath sites and closes the self-consistency directly on the real-frequency axis, which we demonstrate for multiorbital single-site DMFT and for the two-orbital dimer of the two-plane Bethe lattice. The ground state is obtained by the single-site density matrix renormalization group with controlled bond expansion, and the Green's functions by the one-site time-dependent variational principle with the same expansion along a complex-time contour. The impurity tensors of the fork are therefore never updated at two-site cost. The cluster extension rests on three further ingredients. The bath is discretized directly from the matrix-valued hybridization function, with a common energy for each set of bath sites~\cite{FlorezAblan2025,deVega2015}, which avoids the fitting that small baths require. The complex-time Green's function of the cluster is represented by a matrix-valued exponential fit that shares the poles among all its elements~\cite{Zhang2024}. Finally, the positive semidefiniteness of the spectral function is enforced at each frequency, which is essential for the stability of the self-consistency. We validate the solver against the complex-time results of Yu \textit{et al.}~\cite{Yu2026} for two- and three-orbital single-site DMFT, and we apply it to the two-orbital dimer studied by Harland \textit{et al.}~\cite{Harland2019} with continuous-time quantum Monte Carlo, where our zero-temperature spectra show the main spectral features at positions similar to those of the finite-temperature results and resolve them as distinct peaks.

The remainder of this paper is organized as follows. \Cref{sec:cdmft} introduces the two-plane Bethe lattice model, its mapping onto the cluster impurity problem, the bath discretization, and the CDMFT self-consistency. \Cref{sec:ftn} describes the fork tensor network solver with controlled bond expansion, \cref{sec:complexgreensftn} the complex-time Green's functions, and \cref{sec:matrixesprit} the matrix-valued exponential fit and the reconstruction of the real-frequency Green's function. \cref{sec:results} presents the benchmarks for the two-orbital single-site model and the results for the two-orbital dimer, and \cref{sec:conclusion} summarizes our conclusions. The three-orbital benchmark is given in Appendix~\ref{app:3orbsiam}.

\section{Methods}
\subsection{Cluster dynamical mean-field theory}\label{sec:cdmft}

\begin{figure}[htbp]
    \centering
    \includegraphics[width=0.8\linewidth]{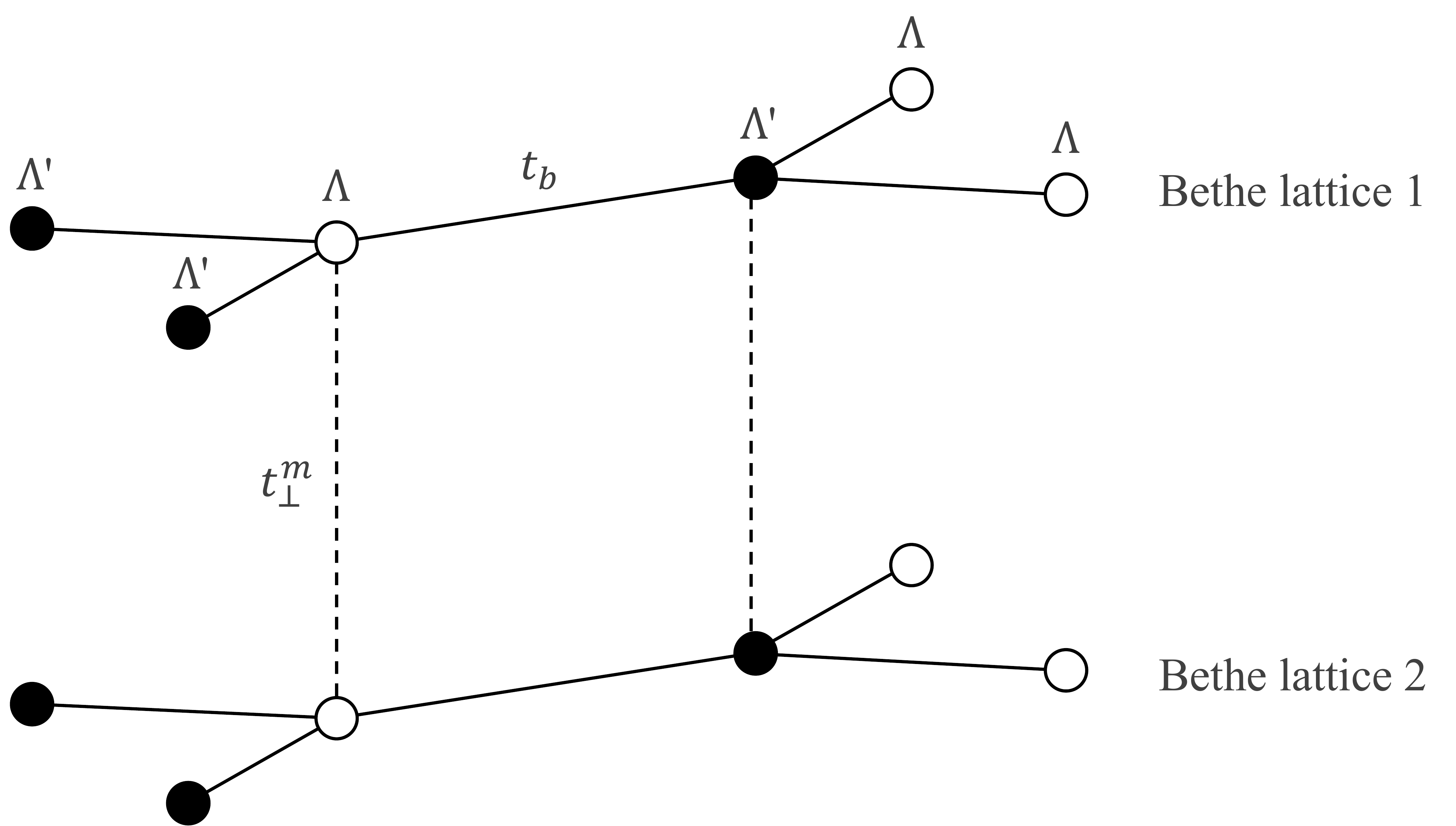}
    \caption{Two-plane Bethe lattice with coordination number $z=3$. The two planes are Bethe lattices with hopping $t_b$, and the dashed lines connect the two sites of a dimer, which are coupled by the intra-dimer hopping $t^{m}_{\perp}$. Open and filled circles mark the two sublattices $\Lambda$ and $\Lambda'$.}
    \label{fig:twoplanebethelattice}
\end{figure}

We consider the Hubbard-Kanamori model on the two-plane Bethe lattice of Ref.~\cite{Harland2019}, shown in \cref{fig:twoplanebethelattice}, and solve it in the limit of infinite coordination number $z\to\infty$ with real-frequency CDMFT at zero temperature. The resulting cluster impurity problem is solved with the complex-time fork tensor network (FTN) impurity solver described in \cref{sec:ftn,sec:complexgreensftn,sec:matrixesprit}.

The lattice Hamiltonian consists of the hopping between dimers and the dimer Hamiltonian, $\hat{H}_{\mathrm{latt}}=\hat{H}_{\mathrm{hop}}+\sum_{\lambda}\hat{H}^{\lambda}_{\mathrm{dimer}}$, with
\begin{equation}
  \hat{H}_{\mathrm{hop}}
  = -t_{b} \sum_{\langle \lambda\lambda' \rangle} \sum_{jm\sigma}
    \left( \hat{a}^{\dagger}_{\lambda jm\sigma}\, \hat{a}_{\lambda' jm\sigma} + \mathrm{H.c.} \right),
  \label{eq:Hhop}
\end{equation}
where $\lambda$ labels the dimers, $\langle\lambda\lambda'\rangle$ denotes pairs of neighboring dimers, and $\sigma\in\{\uparrow,\downarrow\}$, $m\in\{c,d\}$, and $j\in\{1,2\}$ denote the spin, the orbital, and the site within a dimer, respectively; the hopping between neighboring dimers preserves all three indices. The dimer Hamiltonian has the same form for every dimer, so that we omit the index $\lambda$. It comprises the on-site energies, the intra-dimer hopping, and the Kanamori interaction,
\begin{align}
  \begin{split}
  \hat{H}_{\mathrm{dimer}} &= \sum_{jm\sigma} \epsilon_{jm}\, \hat{n}_{jm\sigma} + \hat{H}_{\mathrm{int}} \\
   &\quad -\sum_{m\sigma} t^{m}_{\perp}
     \left( \hat{a}^{\dagger}_{1m\sigma}\hat{a}_{2m\sigma} + \mathrm{H.c.} \right),
  \end{split} \label{eq:Hdimer}\\
  \begin{split}
  \hat{H}_{\mathrm{int}} &= \sum_{j} \Big[\, U \sum_{m} \hat{n}_{jm\uparrow}\hat{n}_{jm\downarrow} \\
   &\quad + \sum_{m<m'} \sum_{\sigma\sigma'} \left( U' - J \delta_{\sigma\sigma'} \right)
       \hat{n}_{jm\sigma}\hat{n}_{jm'\sigma'} \\
   &\quad + J \sum_{m<m'} \left( \hat{a}^{\dagger}_{jm\uparrow}\hat{a}_{jm\downarrow}
       \hat{a}_{jm'\uparrow}\hat{a}^{\dagger}_{jm'\downarrow} + \mathrm{H.c.} \right) \\
   &\quad - J \sum_{m<m'} \left( \hat{a}^{\dagger}_{jm\uparrow}\hat{a}^{\dagger}_{jm\downarrow}
       \hat{a}_{jm'\uparrow}\hat{a}_{jm'\downarrow} + \mathrm{H.c.} \right) \Big],
  \end{split} \label{eq:Hint}
\end{align}
where $\hat{a}^{\dagger}_{jm\sigma}$ ($\hat{a}_{jm\sigma}$) creates (annihilates) an electron and $\hat{n}_{jm\sigma}=\hat{a}^{\dagger}_{jm\sigma}\hat{a}_{jm\sigma}$ is the number operator. In the Kanamori interaction, $U$ is the intra-orbital Coulomb repulsion, $J$ is the Hund's coupling, and the inter-orbital repulsion is $U'=U-2J$, as required by rotational invariance. The on-site energies contain only the chemical potential $\mu$, $\epsilon_{jm}=-\mu$, since, as in Ref.~\cite{Harland2019}, no crystal-field splitting between the $c$ and $d$ orbitals is introduced. In the basis $(1c,2c,1d,2d)$ of site and orbital indices, the intra-dimer hopping is represented by the $4\times4$ matrix $\bm{t}_{\mathrm{loc}}$, whose only non-zero elements, $-t^{c}_{\perp}$ and $-t^{d}_{\perp}$, connect the two sites of each orbital.

In the limit $z\to\infty$, CDMFT maps the lattice model onto a cluster impurity model, in which one dimer, the cluster, is coupled to a non-interacting bath~\cite{Georges1996,Maier2005}. Its Hamiltonian is $\hat{H}_{\mathrm{imp}}=\hat{H}_{\mathrm{dimer}}+\hat{H}_{\mathrm{bath}}+\hat{H}_{\mathrm{hyb}}$ with
\begin{equation}
  \begin{split}
  \hat{H}_{\mathrm{bath}}+\hat{H}_{\mathrm{hyb}} &= \sum_{km\sigma}
     \epsilon_{mk}\, \hat{b}^{\dagger}_{mk\sigma} \hat{b}_{mk\sigma} \\
   &\quad + \sum_{k j m \sigma}
     \left( V_{jmk}\, \hat{b}^{\dagger}_{mk\sigma} \hat{a}_{jm\sigma} + \mathrm{H.c.} \right),
  \end{split}
  \label{eq:Himp}
\end{equation}
where $\hat{b}^{\dagger}_{mk\sigma}$ ($\hat{b}_{mk\sigma}$) creates (annihilates) an electron on the bath site $k\in\{1,\dots,N_b\}$ of orbital $m$ and spin $\sigma$. In the solutions with broken spin symmetry considered below, $\epsilon_{mk}$ and $V_{jmk}$ also depend on $\sigma$, which we suppress. We denote the number of cluster sites by $N_c$ and the number of orbitals by $N_{\mathrm{orb}}$; here $N_c=2$ and $N_{\mathrm{orb}}=2$.

The bath parameters are determined by the hybridization function, which for each $(m,\sigma)$ is an $N_c\times N_c$ matrix in the cluster-site index. Since the FTN solver accommodates hundreds of bath sites, no fit is required: following the parametrization of Ref.~\cite{FlorezAblan2025}, the bath sites are grouped into sets of $N_c$ sites with a common energy, and each set reproduces the weight of $\tilde{\bm{\Delta}}(\omega)=-\frac{1}{2\pi i}\big[\bm{\Delta}(\omega)-\bm{\Delta}^{\dagger}(\omega)\big]$ over one frequency interval exactly (Appendix~\ref{app:disc}).

To allow for antiferromagnetic order, we divide the Bethe lattice into two equivalent sublattices $\Lambda$ and $\Lambda'$~\cite{Harland2019}, see \cref{fig:twoplanebethelattice}; the self-consistency equation then describes a particle on $\Lambda$ that fluctuates through its environment on $\Lambda'$. For a solution with broken spin symmetry, the Weiss field is constructed with the antiferromagnetic condition, $\bm{\mathcal{G}}^{-1}_{\sigma}(\omega)=(\omega+i\eta+\mu)\,\mathbbm{1}-\bm{t}_{\mathrm{loc}}-t_{b}^{2}\,\bm{G}^{\mathrm{loc}}_{-\sigma}(\omega)$, where $\bm{G}^{\mathrm{loc}}_{\sigma}(\omega)$ is the local Green's function, $\mathbbm{1}$ is the unit matrix, and, for each spin $\sigma$, all quantities are $4\times 4$ matrices in the basis $(1c,2c,1d,2d)$. The self-energy follows from the Dyson equation, $\bm{\Sigma}_{\sigma}(\omega)=\bm{\mathcal{G}}^{-1}_{\sigma}(\omega)-[\bm{G}^{\mathrm{imp}}_{\sigma}(\omega)]^{-1}$, and the self-consistency equation for the local Green's function reads
\begin{equation}
  \begin{split}
  \left[ \bm{G}^{\mathrm{loc}}_{\sigma}(\omega) \right]^{-1}
  &= (\omega+i\eta+\mu)\,\mathbbm{1} - \bm{t}_{\mathrm{loc}} \\
  &\quad - t_{b}^{2}\, \bm{G}^{\mathrm{loc}}_{-\sigma}(\omega)
    - \bm{\Sigma}_{\sigma}(\omega).
  \end{split}
  \label{eq:selfcons}
\end{equation}
On the Bethe lattice with $z\to\infty$, the hybridization function follows directly from the local Green's function, $\bm{\Delta}_{\sigma}(\omega)=t_b^{2}\,\bm{G}^{\mathrm{loc}}_{-\sigma}(\omega)$, and its discretization, \cref{eq:Deltadisc}, provides the bath parameters of the next iteration.

Since \cref{eq:selfcons} contains the chemical potential $\mu$, which is fixed by the target occupation of the cluster, it is solved in two nested loops for $\bm{G}^{\mathrm{loc}}$ and $\mu$ at fixed self-energy. The new hybridization function is mixed with the previous one before the next iteration, a weak magnetic field is added to the cluster during the first iterations to reach the solution with broken spin symmetry, and the first iteration starts from the non-interacting cluster; these steps are described in Appendix~\ref{app:disc}. For $N_c=1$, the two-plane Bethe lattice reduces to a single Bethe lattice, the model becomes the multiorbital Hubbard model, and all matrices become scalars for each $(m,\sigma)$.

\subsection{Fork tensor network}\label{sec:ftn}

The computational bottleneck of the CDMFT loop is the impurity solver, which takes the bath parameters $\{\epsilon_{mk},V_{jmk}\}$ and returns the impurity Green's function $\bm{G}^{\mathrm{imp}}(\omega)$. 
\begin{figure}[htbp]
    \centering
    \includegraphics[width=1.0\linewidth]{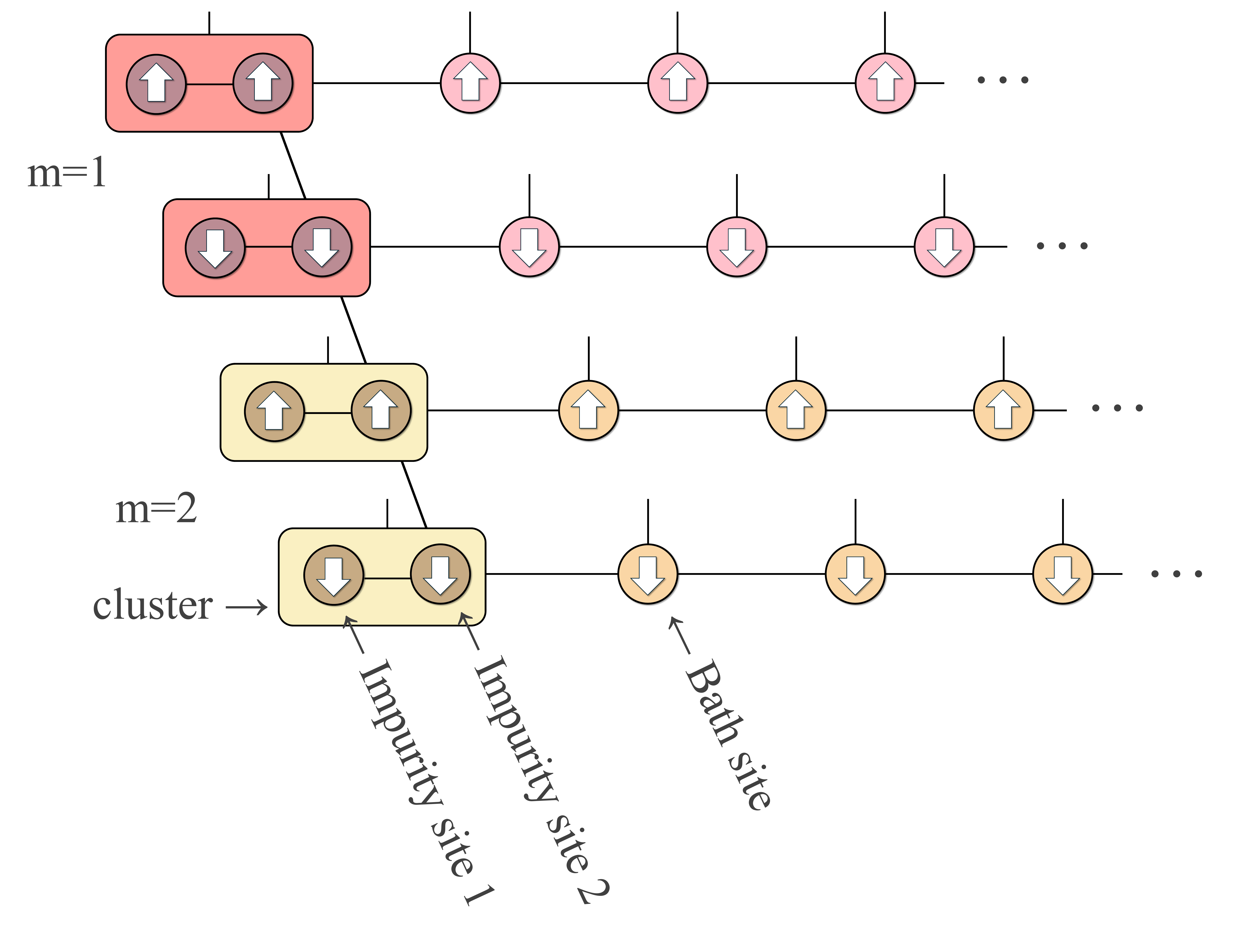}
    \caption{Fork tensor network for the two-orbital, two-site cluster impurity model. Each $(m,\sigma)$ forms one arm. The two arms of $m=1$ (red) and $m=2$ (yellow) are shown. The first tensor of an arm (box) carries the two cluster sites and is connected to the neighboring arms along the backbone; the remaining tensors (circles) carry the bath sites of the same $(m,\sigma)$.}
    \label{fig:clusterftn}
\end{figure}
In $\hat{H}_{\mathrm{imp}}$, the bath sites do not interact with each other, the Kanamori interaction acts only within the cluster, and the hybridization couples a bath site only to the cluster sites of the same $(m,\sigma)$. The FTN~\cite{Bauernfeind2017} reflects this structure (\cref{fig:clusterftn}): each $(m,\sigma)$ forms one arm, whose first tensor carries the $N_c$ cluster sites and whose remaining tensors carry the $N_b$ bath sites hybridizing with them, and the first tensors of all arms are connected along the backbone. We denote the bond dimensions along the backbone and along the arms by $\chi_{x}$ and $\chi_{y}$, respectively. The backbone tensor of each $(m,\sigma)$ has local dimension $2^{N_c}$, so that the intra-dimer hopping acts within a single tensor. Since the FTN contains no loops, a canonical form with respect to any tensor can be constructed as for matrix product states~\cite{Bauernfeind2017}; for $N_c=1$, the network reduces to the fork tensor product state of the single-impurity model.

Within an arm, the bath is represented either in the star geometry of \cref{eq:Himp}, in which every bath site couples directly to the cluster, or in the double-chain geometry~\cite{Kohn2021}, in which the occupied and the empty bath sites are mapped onto two chains interleaved along the arm, with couplings up to next-nearest neighbors; since occupied and empty states are not mixed, the entanglement along the arm remains low. We use the double-chain geometry for the single-impurity calculations. For the cluster, each bath site couples to several cluster sites, so that this mapping does not carry over directly, and we use the star geometry.

The ground state entering the Green's function is obtained by the density matrix renormalization group (DMRG)~\cite{White1992,Schollwoeck2011}, and the time evolution by the time-dependent variational principle (TDVP)~\cite{Haegeman2011,Haegeman2016,Bauernfeind2020}, both formulated on the FTN. Since the cluster impurity Hamiltonian $\hat{H}_{\mathrm{imp}}$ has the $\mathrm{U}(1)\times \mathrm{U}(1)$ symmetry, all tensors conserve the total particle number $N$ and the z component of the total spin, $S_z$, and the ground state is computed in a fixed symmetry sector $(N,S_z)$.
In both, the bond dimension can grow only through two-site updates, which merge and split two neighboring tensors. For a matrix product state, a two-site update is more expensive than a single-site update by a factor of order the local dimension~\cite{Gleis2023}. On the fork, every backbone tensor carries the local dimension $2^{N_c}$, and those in the interior of the backbone carry three virtual legs, so that, with $\chi_x=\chi_y=\chi$, a single-site update of such a tensor scales as $\chi^{4}$~\cite{Bauernfeind2017}, whereas a two-site update of two such tensors scales at least as $\chi^{5}$~\cite{Grundner2025}. Since the backbone tensors are also the most entangled part of the network, their updates dominate the total cost~\cite{Grundner2025}, and the two-site scheme becomes prohibitive for large $\chi$.

We therefore use controlled bond expansion (CBE)~\cite{Gleis2023,Li2024}, which avoids two-site updates altogether. Before each single-site update, the bond towards the next tensor of the sweep is enlarged by those parts of the two-site space that carry significant weight in the action of the Hamiltonian on the state; they are selected at single-site cost by the shrewd selection of Ref.~\cite{Gleis2023}, and the expanded bond is truncated back to $\chi$ when the orthogonality center is moved on. The selection involves only the two tensors adjacent to the bond and their environments, so that it carries over to the fork without modification (Appendix~\ref{app:cbe}). The two-site update of two backbone tensors is thereby replaced by two single-site updates on expanded bonds, which reduces the cost of the dominant step from $\chi^{5}$ to $\chi^{4}$, a power of the bond dimension rather than a constant factor. We use CBE in the single-site DMRG ground-state search and in the single-site TDVP~\cite{Li2024}, where the bond is expanded before the forward evolution of each tensor so that the bond dimension follows the entanglement generated during the time evolution. For the two-site cluster model, the single-site TDVP with CBE is faster than the two-site TDVP by almost an order of magnitude at $\chi=80$, and the difference grows with the bond dimension (Appendix~\ref{app:cbe}).

\subsection{Complex-time Green's function}\label{sec:complexgreensftn}

Tensor-network impurity solvers commonly compute the real-time Green's function and obtain the retarded Green's function on the real-frequency axis by a Fourier transform~\cite{Ganahl2015,Bauernfeind2017}. The accessible time, and with it the frequency resolution, is limited by the growth of entanglement with time, which requires an increasing bond dimension~\cite{Grundner2024}. To overcome this limitation, we perform the time evolution along a contour in the complex-time plane~\cite{Cao2024,Grundner2024,Yu2026} and define the greater and lesser Green's functions as
\begin{align}
  G^{>}_{jj'}(t;\theta) &= -i\,\langle g|\,\hat{a}_{j}\,
     e^{-i(\hat{H}-E_{g})\,e^{-i\theta}t}\,\hat{a}^{\dagger}_{j'}\,|g\rangle,
  \label{eq:Ggreater}\\
  G^{<}_{jj'}(t;\theta) &= i\,\langle g|\,\hat{a}^{\dagger}_{j'}\,
     e^{\,i(\hat{H}-E_{g})\,e^{\,i\theta}t}\,\hat{a}_{j}\,|g\rangle,
  \label{eq:Glesser}
\end{align}
with $t\geq 0$ and $\theta\in[0,\pi/2]$, where $|g\rangle$ is the ground state of $\hat{H}$ with energy $E_g$, and $j,j'=1,\dots,N_c$ label the cluster sites; the Green's functions are diagonal in the arm index $(m,\sigma)$, which we suppress. The angle $\theta=0$ corresponds to real time and $\theta=\pi/2$ to imaginary time.

Since $\hat{H}$ includes the chemical potential and $|g\rangle$ is its ground state, every eigenstate $|n\rangle$ in the sectors with $N\pm1$ particles has a non-negative excitation energy $\Delta_n=E_n-E_g\geq 0$, and its contribution to \cref{eq:Ggreater,eq:Glesser} oscillates as $e^{\mp i\Delta_n\cos\theta\,t}$ and is damped as $e^{-\Delta_n\sin\theta\,t}$. The contributions of high-energy excitations, which drive the fast growth of entanglement, are therefore suppressed; the entanglement grows much more slowly than along the real-time axis, and longer times become accessible at fixed $\chi$. The two contours $e^{\mp i\theta}t$ are complex conjugates of each other, so that both the greater and the lesser Green's function decay for $t\to\infty$. The angle $\theta$ controls a trade-off: a larger $\theta$ reduces the entanglement further, but the reconstruction of the real-frequency spectral function becomes more ill-conditioned, as in the imaginary-time limit $\theta=\pi/2$. We therefore use a small angle, for which the damping suffices to control the entanglement. The complex-time Green's functions are the real-time Green's functions evaluated at the complex times $e^{\mp i\theta}t$; the real-time functions are recovered by rotating the time argument back, which is carried out within the exponential fit of \cref{sec:matrixesprit}.

To evaluate \cref{eq:Ggreater,eq:Glesser}, we obtain $|g\rangle$ and $E_g$ by single-site DMRG with CBE, apply $\hat{a}^{\dagger}_{j'}$ ($\hat{a}_{j'}$) to the backbone tensor of $|g\rangle$, propagate the resulting state along the contour by single-site TDVP with CBE, and take the overlap with $\hat{a}^{\dagger}_{j}|g\rangle$ ($\hat{a}_{j}|g\rangle$). For each $(m,\sigma)$, $N_c$ time evolutions thus yield the full $N_c\times N_c$ matrix of each Green's function; the time evolutions for different arms, cluster sites, and Green's functions are independent of each other and are carried out in parallel (Appendix~\ref{app:cbe}).

\subsection{Matrix-valued ESPRIT}\label{sec:matrixesprit}

The CDMFT self-consistency requires the retarded impurity Green's function $\bm{G}^{\mathrm{imp}}(\omega)$ on the real-frequency axis, whereas the impurity solver yields the greater and lesser Green's functions $G^{\gtrless}_{jj'}(t;\theta)$ on the complex-time contour. The step from the complex-time data to the real-frequency Green's function is analogous to the analytic continuation from imaginary time ($\theta=\pi/2$) and is in general ill-conditioned; it becomes better conditioned as $\theta$ decreases. Following Ref.~\cite{Yu2026}, we represent each complex-time Green's function as a sum of complex exponentials. For each $(m,\sigma)$, the $N_c\times N_c$ Green's function matrix is written as
\begin{equation}
  G^{\gtrless}_{jj'}(t;\theta) \approx \sum_{l=1}^{M^{\gtrless}} R^{\gtrless}_{jj',l}\, e^{s^{\gtrless}_{l} t},
  \label{eq:ansatz}
\end{equation}
where all matrix elements share the exponents $s^{\gtrless}_{l}$ and differ only in the weights $R^{\gtrless}_{jj',l}$, so that the weight of pole $l$ is the $N_c\times N_c$ matrix $\bm{R}^{\gtrless}_{l}$. Since the greater and the lesser Green's function describe different excitations, their numbers of poles $M^{\gtrless}$ and exponents are determined separately. Extracting the parameters $\{M^{\gtrless},\bm{R}^{\gtrless}_{l},s^{\gtrless}_{l}\}$ from the data is a standard signal-processing problem, for which we employ the matrix-valued estimation of signal parameters via rotational invariance techniques (ESPRIT)~\cite{Zhang2024}: the matrices sampled on the time grid are flattened into vectors of length $N_c^{2}$ and arranged into a Hankel matrix, whose singular value decomposition yields the exponents, and the number of poles is set by a tolerance $\epsilon$ on the relative singular values. Applying the scalar-valued ESPRIT to each matrix element separately would yield different numbers and positions of poles for different elements, in conflict with the pole representation of a matrix-valued Green's function, in which all elements share the poles and differ only in their matrix-valued weights~\cite{Zhang2024}. For the single-impurity model ($N_c=1$), the Green's function of each $(m,\sigma)$ is a scalar, and the scalar-valued ESPRIT suffices. The algorithm is summarized in Appendix~\ref{app:matesprit}.

In the representation of \cref{eq:ansatz}, the rotation back to real time, $t\to e^{\pm i\theta}t$, acts only on the exponents and leaves the matrix weights unchanged. The retarded Green's function $G^{R}_{jj'}(t)=\Theta(t)\big[G^{>}_{jj'}(t)-G^{<}_{jj'}(t)\big]$ thus follows in closed form, and, following Ref.~\cite{Yu2026}, a second application of the matrix-valued ESPRIT to $G^{R}(t)$ yields the parameters of its complex pole representation~\cite{Zhang2024,Yu2026},
\begin{equation}
  G^{R}_{jj'}(\omega)\approx\sum_{l=1}^{M}\frac{A_{jj',l}}{\omega+i\eta-\xi_{l}},
  \label{eq:pole_rep}
\end{equation}
where the $M$ poles $\xi_l$ lie in the lower half of the complex frequency plane and $A_{jj',l}$ are the corresponding matrix-valued weights; evaluated on the real-frequency axis, \cref{eq:pole_rep} gives the retarded impurity Green's function $\bm{G}^{\mathrm{imp}}(\omega)$. Exponents that would grow after the rotation to real time and poles with $\mathrm{Im}\,\xi_l>0$, which violate causality, are removed, and the weights of the remaining poles are determined anew~\cite{Yu2026}; this procedure is given in Appendix~\ref{app:esprit}.

For each $(m,\sigma)$, the impurity spectral function $\bm{A}^{\mathrm{imp}}(\omega)=-\frac{1}{2\pi i}\big[\bm{G}^{\mathrm{imp}}(\omega)-\bm{G}^{\mathrm{imp}}(\omega)^{\dagger}\big]$ is an $N_c\times N_c$ Hermitian matrix that must be positive semidefinite at every frequency; in the scalar case, this condition reduces to $A(\omega)\geq 0$. If the fitted $\bm{A}^{\mathrm{imp}}$ violates this condition at some frequency, we observe that an unphysical peak grows there in the course of the CDMFT iterations of \cref{eq:selfcons}. We therefore project $\bm{A}^{\mathrm{imp}}$ onto its positive semidefinite part at each frequency, $\bm{A}^{\mathrm{PSD}}=\sum_{n}\max[\lambda_n,0]\,\bm{u}_n\bm{u}_n^{\dagger}$, where $\lambda_n$ and $\bm{u}_n$ are the eigenvalues and eigenvectors of the Hermitian part $(\bm{A}^{\mathrm{imp}}+\bm{A}^{\mathrm{imp}\,\dagger})/2$. This projection is essential for the stability of the self-consistency. Since it removes negative eigenvalues and thus adds spectral weight, which violates the sum rule $\int\bm{A}(\omega)\,d\omega=\mathbbm{1}$, the tolerance $\epsilon$ of the matrix-valued ESPRIT is chosen such that the projection is needed as little as possible. Since the Dyson equation requires $\bm{G}^{\mathrm{imp}}$ rather than $\bm{A}^{\mathrm{imp}}$, the impurity Green's function is reconstructed from the projected spectral function,
\begin{equation}
  \bm G^{\mathrm{imp}}(\omega)=\mathcal P\!\int d\omega'\,\frac{\bm A^{\mathrm{PSD}}(\omega')}{\omega-\omega'}-i\pi\bm A^{\mathrm{PSD}}(\omega),
  \label{eq:GfromA}
\end{equation}
where the Hermitian part is the principal-value Hilbert transform of $\bm{A}^{\mathrm{PSD}}$~\cite{Georges1996} and the anti-Hermitian part is $-i\pi\bm{A}^{\mathrm{PSD}}$.

\section{results}\label{sec:results}

We apply the solver to three models. For the two-orbital and three-orbital single-impurity Anderson models, we compare the spectral functions with those of Figs.~7 and 8 of Ref.~\cite{Yu2026} for the same parameters.
The three-orbital results are presented in Appendix~\ref{app:3orbsiam}. These calculations use the CDMFT solver in the single-impurity case ($N_c=1$) and serve as a benchmark before we turn to the cluster. We then perform CDMFT calculations for the two-orbital, two-site cluster impurity model of the two-plane Bethe lattice.

In the single-impurity calculations, all energies are given in units of the half bandwidth $D$ of the semicircular non-interacting density of states $\rho_0(\omega)=(2/\pi D^2)\sqrt{D^2-\omega^2}$. Since the hybridization function is updated directly from the impurity Green's function, $\Delta(\omega)=t_b^{2}G^{\mathrm{imp}}(\omega)$ with $t_b=D/2$ (\cref{sec:cdmft}), and the chemical potential $\mu$ is fixed to the value that reproduces the filling of Ref.~\cite{Yu2026}, neither the inner nor the outer loop of Appendix~\ref{app:disc} is needed. The two-orbital model at half filling has orbital, spin, and particle-hole symmetry, so that only the greater Green's function of a single $(m,\sigma)$ needs to be computed by time evolution; the lesser Green's function follows from particle-hole symmetry as $G^<(z^*)=[G^>(z)]^*$. The three-orbital model lacks particle-hole symmetry, and both the greater and the lesser Green's function are computed for a single $(m,\sigma)$. Since the Green's function of each $(m,\sigma)$ is a scalar, the scalar-valued ESPRIT is used.

\subsection{Two-orbital single-impurity Anderson model}\label{sec:2orbsiam}

\begin{figure}[htbp]
    \centering
    \includegraphics[width=0.9\linewidth]{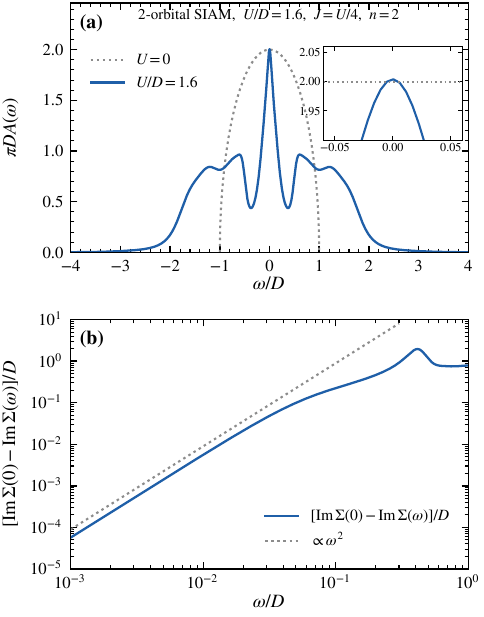}
    \caption{Spectral function and self-energy of the two-orbital single-impurity Anderson model at half filling ($n=2$) with $U/D=1.6$ and $J=U/4$. (a) Spectral function $\pi DA(\omega)$ of a single $(m,\sigma)$ (solid line) and semicircular density of states of the non-interacting Bethe lattice (dotted line). The inset enlarges the region around $\omega=0$; the dotted line marks the non-interacting value $\pi DA(0)=2$. (b) Low-frequency behavior of the self-energy. The solid line shows $[\mathrm{Im}\,\Sigma(0)-\mathrm{Im}\,\Sigma(\omega)]/D$ on a log-log scale, and the dotted line is proportional to $\omega^2$. The peak at $\omega/D\simeq 0.41$ lies close to the minimum of the spectral function in (a).}
    \label{fig:2orbsaim}
\end{figure}

We solve the two-orbital Anderson impurity model at half filling ($n=2$) with $U/D = 1.6$, $J = U/4$, and half bandwidth $D=2$. The bath is represented in the double-chain geometry with $N_b=99$ bath sites per $(m,\sigma)$, and the bond dimension is $\chi=40$. The complex-time evolution uses the angle $\theta=0.3$ and the time step $D\,\Delta t=0.1$ up to $D\,t_{\max}=50$, the broadening $\eta=10^{-10}$, and the ESPRIT tolerance is $\epsilon=10^{-6}$. These settings are identical to those of Fig.~7 of Ref.~\cite{Yu2026}. 

\cref{fig:2orbsaim}(a) shows the spectral function $\pi DA(\omega)$. Since the orbitals and spins are degenerate, the spectral function of a single $(m,\sigma)$ is shown, together with the semicircular density of states of the non-interacting Bethe lattice (dotted line). A peak appears at $\omega=0$, Hubbard bands form at $|\omega/D|\gtrsim1$, and an additional peak emerges between them at $|\omega/D|\simeq0.6$. This structure is consistent with the half-filling result of Fig.~7 of Ref.~\cite{Yu2026}. At $\omega=0$, the spectral function takes the value $\pi DA(0)=2.0042$, close to the non-interacting value, and the sum rule $\int A(\omega)\,d\omega=1$ is fulfilled to about $10^{-4}$.

\cref{fig:2orbsaim}(b) shows $[\mathrm{Im}\,\Sigma(0)-\mathrm{Im}\,\Sigma(\omega)]/D$ on a log-log scale to examine the low-frequency behavior of $-\mathrm{Im}\,\Sigma(\omega)$. At $\omega=0$, the imaginary part is small, $\mathrm{Im}\,\Sigma(0)\simeq 0.002D$, and $\mathrm{Re}\,\Sigma(0)=1.4D$. At low frequencies, $-\mathrm{Im}\,\Sigma(\omega)$ follows the Fermi-liquid form proportional to $\omega^2$~\cite{Yu2026}. The peak at $\omega/D\simeq 0.41$ lies close to the minimum of the spectral function at $\omega/D\simeq 0.37$ in \cref{fig:2orbsaim}(a).

\subsection{Two-orbital two-site cluster impurity model}\label{sec:2impcAIM}

\begin{figure}[htbp]
    \centering
    \includegraphics[width=1\linewidth]{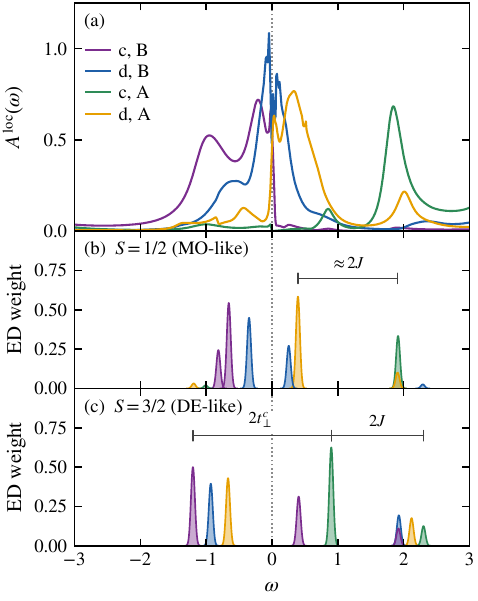}
    \caption{Local spectral function of the two-orbital dimer compared with exact diagonalization (ED) of the isolated dimer, for $U=4.5$, $J=0.7$, $t^{c}_{\perp} = 1.05$, $t^{d}_{\perp} = 0.2$, $t_b = 0.34$, and $N = 3$ electrons in the cluster.
    (a) CDMFT result in the bonding/antibonding basis, $\tilde{A}_{m,B/A}(\omega) = [T A^{\mathrm{loc}}_{m}(\omega) T^{T}]_{B/A,B/A}$ with the transformation $T$ defined in the text, averaged over spin. The four curves show the bonding ($B$) and antibonding ($A$) components of the $c$ and $d$ orbitals.
    (b) ED weights of the poles for excitations from the molecular-orbital (MO)-like multiplet with $S=1/2$, which lies 11.4~meV above the ground state of the isolated dimer.
    (c) Same as (b) for the double-exchange (DE)-like multiplet with $S=3/2$, which is the ground state of the isolated dimer. The height of each peak gives the weight of the pole. The colors are the same as in (a). The horizontal bars mark the energy separations $2t^c_{\perp}$ and $2J$, and the vertical dotted line marks the Fermi level $\omega=0$.
    } 
    \label{fig:AlocBA}
\end{figure}

We now apply CDMFT to the two-plane Hubbard-Kanamori model of Ref.~\cite{Harland2019} introduced in \cref{sec:cdmft}, with two orbitals, $c$ and $d$, per site. All energies are given in units of eV, and the unit is omitted below. The interaction parameters are $U = 4.5$ and $J = 0.7$ ($U' = U - 2J = 3.1$), the intra-dimer hopping amplitudes are $t^{c}_{\perp} = 1.05$ and $t^{d}_{\perp} = 0.2$, the Bethe lattice hopping is $t_b = 0.34$, and the cluster contains $N = 3$ electrons. These parameters lie in the crossover region identified in Ref.~\cite{Harland2019} at $t^{c}_{\perp} = 1.05$, where approximately 90\% of the ground state at $t_b = 0.35$ consists of the molecular-orbital (MO) state. Since the chemical potential is not specified in Ref.~\cite{Harland2019}, $\mu$ is adjusted by the outer loop to give $N = 3$, which yields $\mu \approx 2.55$.

The impurity solver uses the star geometry with $N_b = 120$ bath sites per $(m,\sigma)$, and the time evolution is carried out for 500 steps with $\chi=120$, $\theta = 0.2$, and $\Delta t = 0.1$. In the self-consistency equation, \cref{eq:selfcons}, we use the broadening $\eta = 10^{-2}$ and the mixing parameter $\gamma=0.3$. The ESPRIT tolerance $\epsilon$ is chosen separately for each arm and ranges between $10^{-4}$ and $10^{-2}$ (\cref{sec:matrixesprit}).

To obtain the antiferromagnetic solution, a weak magnetic field is added to the on-site energies of the cluster during the first iterations, as described in \cref{sec:cdmft}. The field takes the form $(-h, -h, -h, 0)$ for $(m, \sigma) = (c\uparrow, c\downarrow, d\uparrow, d\downarrow)$ on both cluster sites, and the spin indices are interchanged on the opposite sublattice. The calculation is run with $h = 0.02$ for several iterations, after which $h$ is set to zero. \cref{fig:AlocBA}(a) presents the spectral function obtained after several further iterations with $h=0$.

The results of Ref.~\cite{Harland2019} were obtained with the continuous-time quantum Monte Carlo solver in the hybridization expansion (CTHYB) at $T=0.01$, where the impurity Green's function and the self-consistency loop are formulated on the Matsubara axis and the spectral function is obtained afterwards by analytic continuation with the stochastic optimization method~\cite{Mishchenko2000,Krivenko2019}. 
In contrast, our solver computes the impurity Green's function by complex-time evolution at $T=0$ and transforms it to real frequencies (\cref{sec:matrixesprit}), and the self-consistency loop, \cref{eq:selfcons}, is closed directly on the real-frequency axis. A comparison with the spectrum at $t_b = 0.34$ in Fig.~9 of Ref.~\cite{Harland2019} shows similar positions of the peak near the Fermi level and of the structures around $\omega \approx - 1$. In \cref{fig:AlocBA}(a), distinct peaks appear at $\omega \approx -1.0, -0.2, 0.3,$ and $1.9$, whereas the spectrum of Ref.~\cite{Harland2019} is broader. 
This difference may be attributed in part to the analytic continuation, which tends to broaden the peaks, and in part to the finite temperature: at $T=0.01$, the system is an incoherent metal with local moments for $0.33<t_b<0.4$~\cite{Harland2019}. For each $(m,\sigma)$ component, the sum rule $\int A(\omega)\,d\omega= 1$ is fulfilled to about $4\times10^{-3}$.

\cref{fig:AlocBA}(a) shows the spectral function in the bonding/antibonding basis. The transformation 
\begin{align}
 T = \frac{1}{\sqrt{2}} \begin{pmatrix} 1 & 1 \\ 1 & -1 \end{pmatrix}   
\end{align}
from the cluster sites to the bonding ($B$) and antibonding ($A$) orbitals~\cite{Harland2019} gives $\tilde{A} = T A^{\mathrm{loc}} T^T$, whose four diagonal components $(c,B)$, $(c,A)$, $(d,B)$, and $(d,A)$, averaged over spin, are shown; the off-diagonal components are negligible. For the $c$ orbital, the weight of $(c,B)$ lies mainly at $\omega < 0$, while that of $(c,A)$ is centered at $\omega \approx 1.9$. In contrast, both $(d,B)$ and $(d,A)$ have weight near the Fermi level $\omega \approx 0$. The corresponding occupations, obtained by integrating $\tilde{A}_{m, B/A}(\omega)$ up to the Fermi level, are $1.59$ for $(c,B)$, $0.30$ for $(c,A)$, $0.83$ for $(d,B)$, and $0.27$ for $(d,A)$, which sum to $N=2.98$; the small deficit from $3$ reflects the sum rule of the pole representation. The ground-state occupations of the cluster, summed over the two sites, are $n_\uparrow = 0.969$ and $n_\downarrow = 0.957$ for the $c$ orbital and $n_\uparrow = 0.719$ and $n_\downarrow = 0.381$ for the $d$ orbital, which correspond to the magnetic moments $m_c = 0.012$ and $m_d = 0.338$ per dimer. The total ground-state occupation of the cluster is $N=3.03$, close to the target value of $3$.

The ground state of the isolated dimer is the DE-like multiplet with $S=3/2$, and the MO-like multiplet with $S=1/2$ lies 11.4~meV above it. \cref{fig:AlocBA}(b) and \cref{fig:AlocBA}(c) show the poles of the excitations from these two multiplets. Although the ground state of the isolated dimer has $S=3/2$, the hybridization with the bath mixes in the $S=1/2$ component, so that excitations from the $S=1/2$ multiplet additionally appear in the spectral function.

\begin{figure}[htbp]
    \centering
    \includegraphics[width=1\linewidth]{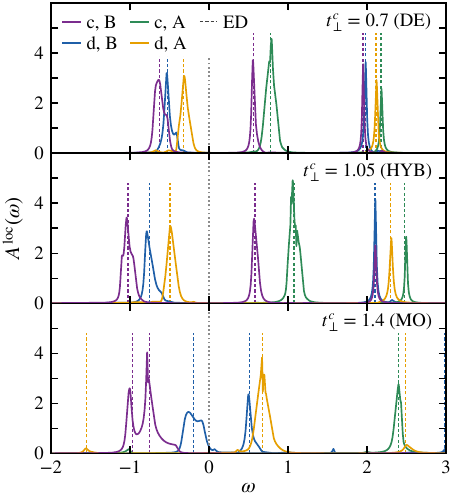}
    \caption{Local spectral function $A^{\mathrm{loc}}(\omega)$ of the dimer in the bonding $(B)$ and antibonding $(A)$ basis, averaged over spin, for $t_b=0.1$, $t^d_{\perp}=0.2$, and $t^c_{\perp}=0.7$ (DE), $1.05$ (HYB), and $1.4$ (MO). The other parameters are the same as in \cref{fig:AlocBA}. Solid curves are our CDMFT results. Vertical dashed lines mark the pole positions obtained by ED of the isolated dimer, in the color of the corresponding curve. The gray dotted line marks $\omega=0$.
    }
    \label{fig:figtb0p1_Aloc}
\end{figure}

We next consider a weak Bethe lattice hopping, $t_b=0.1$, with $t^c_{\perp}=0.7$, $1.05$, and $1.4$, which lie in the double-exchange (DE) regime, in the crossover region (HYB), and in the molecular-orbital (MO) regime of Ref.~\cite{Harland2019}, respectively. The other parameters and the solver settings are the same as above, except that the ESPRIT tolerance $\epsilon$ ranges between $10^{-7}$ and $10^{-2}$. The chemical potential that gives $N=3$ is $\mu \approx 2.32$, $2.37$, and $2.39$, respectively. For $t^c_{\perp}=1.4$, the magnetic field takes the form $(-h, -h, -h, 0)$ given above, while for $t^c_{\perp}=0.7$ and $1.05$ it takes the form $(-h, +h, -h, +h)$. In all three cases, the calculation is run with $h=0.02$ for several iterations, after which $h$ is set to zero. \cref{fig:figtb0p1_Aloc} presents the spectral function in the bonding/antibonding basis obtained after several further iterations with $h=0$. The vertical dashed lines mark the positions of the poles obtained by ED for the ground state of the isolated dimer with the same $U$, $J$, $t^c_{\perp}$, $t^d_{\perp}$, and $\mu$.

In \cref{fig:figtb0p1_Aloc}, a peak of the same component appears at the position of each ED pole for all three values of $t^{c}_{\perp}$, and the peak positions differ from the ED poles by at most $0.064$. The hybridization function, $\bm{\Delta}_{\sigma}=t_b^{2}\,\bm{G}^{\mathrm{loc}}_{-\sigma}$, carries the prefactor $t_b^2$, which at $t_b=0.1$ is smaller than at $t_b=0.34$ by a factor of about 12, so that the hybridization with the bath is weak. The cluster is therefore closer to the atomic limit of Ref.~\cite{Harland2019}, $t_b\to0$. In this limit, the dimers are isolated and the spectral function reduces to the ED poles. The agreement with the ED poles is thus better than at $t_b=0.34$.

\section{conclusion}\label{sec:conclusion}

We have presented a zero-temperature impurity solver for CDMFT that combines controlled bond expansion with complex-time evolution on fork tensor product states. In the fork tensor network, the backbone tensors that carry the cluster sites have up to three virtual legs and dominate the cost of both the ground-state search and the time evolution. CBE replaces the two-site updates of these tensors by single-site updates on expanded bonds in the DMRG and in the TDVP, which lowers the cost of the dominant step by a power of the bond dimension; for the two-site cluster, the time evolution becomes faster by almost an order of magnitude. Complex-time evolution damps the high-energy excitations, so that long times become accessible at moderate bond dimension. The real-frequency Green's function of the cluster is obtained by the matrix-valued ESPRIT, in which all matrix elements share the poles, and the positive semidefinite projection of the spectral function stabilizes the self-consistency. Together with the direct discretization of the matrix-valued hybridization function, which accommodates hundreds of bath sites without a fit, this closes the CDMFT self-consistency directly on the real-frequency axis.

For two- and three-orbital single-site DMFT, the spectral functions agree with the complex-time results of Ref.~\cite{Yu2026}, and the solver reproduces the Fermi-liquid behavior of the self-energy at half filling and the multiplet structure of the three-orbital model (Appendix~\ref{app:3orbsiam}). For the two-orbital dimer of the two-plane Bethe lattice in the crossover region of Ref.~\cite{Harland2019}, we obtained the antiferromagnetic solution at zero temperature. The main spectral features appear at positions similar to those of the finite-temperature CTHYB results of Ref.~\cite{Harland2019}, and our zero-temperature spectra resolve them as distinct peaks. In the bonding/antibonding basis, the $c$ electrons occupy mainly the bonding orbital, consistent with the formation of the molecular-orbital state, whereas the magnetic moment resides in the $d$ orbital.

Since CBE involves only the two tensors adjacent to a bond and their environments, it applies to any tree tensor network, and its gain over two-site updates grows with the number of virtual legs of the updated tensors. Natural extensions of this work include clusters with more sites and orbitals and a generalization of the double-chain geometry to the bath of a cluster. Because the solver yields spectral functions directly on the real-frequency axis without analytic continuation, it is well suited for multiorbital cluster calculations of dimerized transition-metal compounds, whose spectra can be compared with photoemission experiments.

\acknowledgments 
T.-Y.~L. and H.-Y.~L. were supported by the Basic Science Research Program through the National Research Foundation (NRF) of Korea funded by the Ministry of Science and ICT (MSIT) (Grant Nos. RS-2023-00220471 and RS-2025-16064392). H.-S.~K. acknowledges the support of the Basic Science Research Program through the NRF Korea by the MSIT (RS-2025-23525695), and additionally thanks the KENTECH Startup Funding for new faculty members (202500005A). J.-H.~C. was supported by the National Research Foundation of Korea (NRF) grants funded by the Korean government (MSIT: No. RS-2023-00214464, No. RS-2024-00442710).

\appendix

\section{Details of the CDMFT iteration}\label{app:disc}

\subsection{Bath discretization}

\begin{figure}[htbp]
    \centering
    \includegraphics[width=1.0\linewidth]{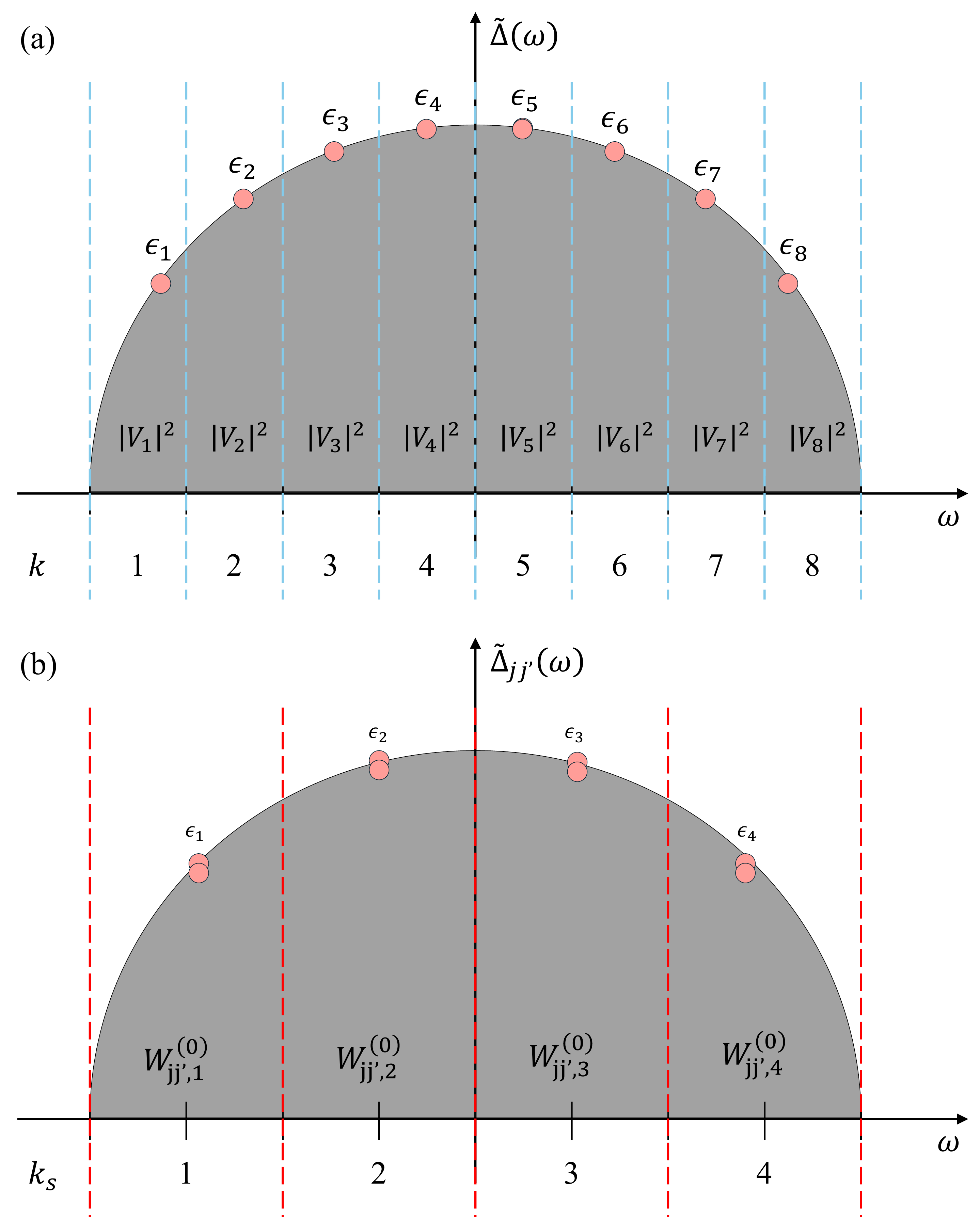}
    \caption{Schematic of the bath discretization for a semicircular hybridization function with $N_b=8$ bath sites and intervals of equal width. (a) Single site ($N_c=1$): each interval $I_k$ yields one bath site with $|V_k|^2=\int_{I_k}\tilde{\Delta}(\omega)\,d\omega$ and $\epsilon_k$ at the centroid of $\tilde{\Delta}(\omega)$ over $I_k$. (b) Cluster ($N_c=2$): the curve represents a generic element $\tilde{\Delta}_{jj'}(\omega)$; each interval $I_{k_s}$ defines an $N_c\times N_c$ weight matrix $\bm{W}^{(0)}_{k_s}$, \cref{eq:W0W1}, and a single energy $\epsilon_{k_s}$, \cref{eq:eps_ks}, shared by $N_c$ bath sites.
    }
    \label{fig:sdrdisc}
\end{figure}

For each $(m,\sigma)$, we group the $N_b=N_s N_c$ bath sites into $N_s$ sets of $N_c$ sites with a common energy $\epsilon_{k_s}$, so that the discretized hybridization function reads
\begin{equation}
  \Delta^{\mathrm{disc}}_{jj'}(\omega)
  = \sum_{k_{s}, n}
    \frac{V_{j,(k_{s},n)}\, V^{*}_{j',(k_{s},n)}}{\omega+i\eta-\epsilon_{k_{s}}},
  \label{eq:Deltadisc}
\end{equation}
where $k_s$ labels the sets, $n$ the bath sites within a set, and the indices $m$ and $\sigma$ are suppressed. The parameters $\epsilon_{k_s}$ and $V_{j,(k_s,n)}$ are obtained from the hybridization function as follows. We divide the frequency axis into $N_s$ intervals $I_{k_s}$ and define for each interval the weight matrix and its first moment,
\begin{equation}
  \bm{W}^{(0)}_{k_s} = \int_{I_{k_s}} \tilde{\bm{\Delta}}(\epsilon)\, d\epsilon,
  \qquad
  \bm{W}^{(1)}_{k_s} = \int_{I_{k_s}} \epsilon\, \tilde{\bm{\Delta}}(\epsilon)\, d\epsilon,
  \label{eq:W0W1}
\end{equation}
where $\tilde{\bm{\Delta}}(\omega)=-\frac{1}{2\pi i}\big[\bm{\Delta}(\omega)-\bm{\Delta}^\dagger(\omega)\big]$ is Hermitian and positive semidefinite. The energy of each set is fixed by the first moment, and the hybridization amplitudes follow from the eigenvalue decomposition $W^{(0)}_{jj',k_s}=\sum_n \lambda_{n,k_s}\,u_{j,n,k_s}u^{*}_{j',n,k_s}$:
\begin{align}
  \epsilon_{k_s}
  &= \frac{\mathrm{Re}\,\mathrm{Tr}\big[\bm{W}^{(1)}_{k_s}\bm{W}^{(0)}_{k_s}\big]}
          {\mathrm{Tr}\big[\bm{W}^{(0)}_{k_s}\bm{W}^{(0)}_{k_s}\big]},
  \label{eq:eps_ks}\\
  V_{j,(k_s,n)} &= \sqrt{\lambda_{n,k_s}}\; u_{j,n,k_s},
  \label{eq:V_ks}
\end{align}
so that $\sum_n V_{j,(k_s,n)}V^{*}_{j',(k_s,n)}=W^{(0)}_{jj',k_s}$ and the weight of each interval is reproduced exactly. For $N_c=1$, the construction reduces to the standard discretization of a scalar hybridization function~\cite{deVega2015}. \cref{fig:sdrdisc} illustrates the construction for a semicircular hybridization function. As a safeguard, the weight matrices $\bm{W}^{(0)}_{k_s}$ can be projected onto their positive semidefinite part before the eigenvalue decomposition, in the same way as the spectral function in \cref{sec:matrixesprit}.

\subsection{Self-consistency loop}

Since \cref{eq:selfcons} contains the chemical potential $\mu$, which is fixed by the target occupation of the cluster, we solve it in two nested loops. The inner loop iterates \cref{eq:selfcons} at fixed $\mu$ until $\bm{G}^{\mathrm{loc}}$ converges, and the outer loop adjusts $\mu$, common to all $(m,\sigma)$, until the cluster occupation
\begin{equation}
    N(\mu) = -\frac{1}{\pi} \sum_{\sigma} \int^0_{-\infty} d\omega \; \mathrm{Tr}\, \mathrm{Im}\, \bm{G}^{\mathrm{loc}}_{\sigma}(\omega; \mu)
    \label{eq:occ}
\end{equation}
reaches its target value. The nested loops do not call the impurity solver; they form a post-processing step within one CDMFT iteration. One CDMFT iteration thus consists of the following steps. (i) The bath parameters $\{\epsilon_{mk},V_{jmk}\}$ are obtained from the current hybridization function by the discretization of \cref{eq:Deltadisc}. (ii) The impurity solver returns $\bm{G}^{\mathrm{imp}}$. (iii) The self-energy follows from the Dyson equation, and the nested loops yield $\bm{G}^{\mathrm{loc}}$ and $\mu$. (iv) The new hybridization function, $\bm{\Delta}_{\sigma}=t_b^{2}\,\bm{G}^{\mathrm{loc}}_{-\sigma}$, is mixed with the previous one, $\bm{\Delta}\leftarrow\gamma\,\bm{\Delta}^{\mathrm{new}}+(1-\gamma)\,\bm{\Delta}^{\mathrm{old}}$, and used in the next iteration. To reach a solution with broken spin symmetry, a weak magnetic field $h$ is added to the on-site energies of the cluster during the first few iterations.

\subsection{Initial local Green's function}

The first CDMFT iteration starts from the non-interacting cluster. For each $(m,\sigma)$, the imaginary part of the non-interacting local Green's function is given by two semicircular densities of states $\rho_{\pm}$ centered at $\mp t^{m}_{\perp}$,
\begin{equation}
  \mathrm{Im}\, \bm{G}^{0,\mathrm{loc}}(\omega)
  = -\begin{pmatrix}
      \rho_{+}+\rho_{-} & \rho_{+}-\rho_{-} \\
      \rho_{+}-\rho_{-} & \rho_{+}+\rho_{-}
    \end{pmatrix},
  \label{eq:seed}
\end{equation}
where the matrix indices are the cluster sites $j\in\{1,2\}$ and $\rho_{\pm}(\omega)=D^{-2}\sqrt{D^{2}-\left(\omega\pm t^{m}_{\perp}\right)^{2}}$, with $D$ the half bandwidth. The real part follows from the principal-value Hilbert transform of the imaginary part.
Before the first CDMFT iteration, the nested loops of \cref{eq:selfcons} are solved with $\bm{\Sigma}=0$, starting from \cref{eq:seed}. They yield the non-interacting local Green's function $\bm{G}^{0,\mathrm{loc}}$ at the target occupation. In the first CDMFT iteration, $\bm{G}^{0,\mathrm{loc}}$ takes the place of the local Green's function of the previous iteration. In step (i), the hybridization function $\bm{\Delta}_{\sigma}=t_b^{2}\,\bm{G}^{0,\mathrm{loc}}_{-\sigma}$ is discretized into the bath of the first impurity problem, and in step (iii), the self-energy follows from the Dyson equation with the Weiss field constructed from $\bm{G}^{0,\mathrm{loc}}$. The nested loops then yield $\bm{G}^{\mathrm{loc}}$ and $\mu$ of the first iteration, as in every later iteration.

\section{Controlled bond expansion on the fork tensor network}\label{app:cbe}

\subsection{Shrewd selection}

Before each single-site update, the bond towards the next tensor of the sweep is enlarged by those parts of the two-site space, orthogonal to the current single-site space, that carry significant weight in the action of the Hamiltonian on the state. They are found by the shrewd selection of Ref.~\cite{Gleis2023}, a preselection of at most $\chi/w$ states, with $w$ the bond dimension of the Hamiltonian, followed by a truncation to a small fraction $\delta$ of the bond dimension, both at single-site cost. The single-site update is then performed on the expanded bond, which is truncated back to $\chi$ when the orthogonality center is moved to the next tensor. The selection involves only the two tensors adjacent to the bond and their environments; for a bond along the backbone, the environment of a backbone tensor consists of the environments above and below it along the backbone and the environment of its arm. In the single-site TDVP, the bond is expanded before the forward evolution of each tensor~\cite{Li2024}, and the complex time step enters the integrator as $\Delta t\,e^{\mp i\theta}$ along the two contours of \cref{eq:Ggreater,eq:Glesser}.

\subsection{Evaluation of the Green's functions}

The creation (annihilation) operator $\hat{a}^{\dagger}_{j'}$ ($\hat{a}_{j'}$) is applied to the backbone tensor of the ground state $|g\rangle$, the resulting state $\hat{a}^{\dagger}_{j'}|g\rangle$ ($\hat{a}_{j'}|g\rangle$) is propagated along the contour $e^{-i\theta}t$ ($e^{i\theta}t$) by single-site TDVP with CBE, and the Green's function follows from the overlap with $\hat{a}^{\dagger}_{j}|g\rangle$ ($\hat{a}_{j}|g\rangle$). Since only the state with the excitation on site $j'$ is time evolved, $N_c$ time evolutions suffice for the full $N_c\times N_c$ Green's function matrix of each $(m,\sigma)$. Since neither the hopping nor the hybridization mixes different $(m,\sigma)$, the Green's function is block-diagonal in the arm index, and each block is an $N_c\times N_c$ matrix; for $N_c=1$, it is a scalar. In general, $4N_{\mathrm{orb}}N_c$ time evolutions are required, two per $(m,\sigma)$ and cluster site for the greater and the lesser function. Orbital symmetry reduces this number by a factor of $N_{\mathrm{orb}}$, spin symmetry by a factor of two, and particle-hole symmetry by another factor of two, since the lesser function then follows from the greater one. The time evolutions are independent of each other and can be carried out in parallel.

\subsection{Computation time}

\begin{figure}[htbp]
    \centering
    \includegraphics[width=1.0\linewidth]{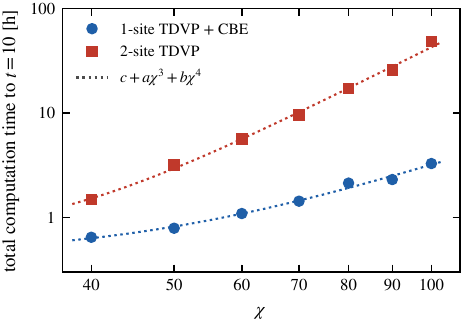}
    \caption{ Total computation time for propagating the greater Green's function $G^{>}(t)$ up to $t=10$ with the single-site TDVP with CBE and with the two-site TDVP, as a function of the bond dimension $\chi$ on a log-log scale.
    Dotted lines are fits to $c+a\chi^3+b\chi^4$, with $c=0.43$, $a=3.5\times10^{-6}$, and $b=-6.8\times10^{-9}$ for the single-site TDVP with CBE and $c=0.65$, $a=-5.4\times10^{-6}$, and $b=4.7\times10^{-7}$ for the two-site TDVP.
    }
    \label{fig:tdvpbenchtime}
\end{figure}

We compare the computation time of the single-site TDVP with CBE and of the two-site TDVP for the two-site cluster model in the star geometry with $N_b=120$ bath sites per $(m,\sigma)$. Both start from the same ground state, obtained in the first CDMFT iteration, and propagate the greater Green's function $G^{>}(t)$ of one cluster site for 100 steps with $\theta=0.2$ and $\Delta t=0.1$, that is, up to $t=10$. All runs use four Julia threads with single-threaded BLAS on an Intel Xeon Platinum 8452Y processor. \cref{fig:tdvpbenchtime} shows the total computation time as a function of the bond dimension $\chi$. The time of the two-site scheme grows much faster with $\chi$, and at $\chi=80$ the single-site scheme with CBE is faster by almost an order of magnitude.

\section{From complex time to real frequency}\label{app:esprit}

Once the complex-time Green's functions are represented as in \cref{eq:ansatz}, the real-time Green's functions follow by rotating the time argument, $t\to e^{\pm i\theta}t$, which rotates only the exponents,
\begin{align}
  G^{>}_{jj'}(t) &\approx \sum_{l=1}^{M^{>}} R^{>}_{jj',l}\,
     e^{\,s^{>}_{l}e^{\,i\theta}t},
  \label{eq:rot_greater}\\
  G^{<}_{jj'}(t) &\approx \sum_{l=1}^{M^{<}} R^{<}_{jj',l}\,
     e^{\,s^{<}_{l}e^{-i\theta}t}.
  \label{eq:rot_lesser}
\end{align}
The retarded Green's function
\begin{equation}
  G^{R}_{jj'}(t)=\Theta(t)\big[G^{>}_{jj'}(t)-G^{<}_{jj'}(t)\big]
  \label{eq:GR_t}
\end{equation}
then follows, and by the residue theorem the pole representation of \cref{eq:pole_rep} corresponds to
\begin{equation}
    G^{R}_{jj'}(t)\approx-i\,\Theta(t)\sum_{l=1}^{M}A_{jj',l}\,e^{-i\xi_{l}t},
    \label{eq:pole_GR_t}
\end{equation}
which has the form of \cref{eq:ansatz} with $\xi_{l}=i\,s_{l}$ and $A_{jj',l}=i\,R_{jj',l}$. The parameters $\{\xi_l,A_{jj',l}\}$ are therefore obtained by applying the matrix-valued ESPRIT (Appendix~\ref{app:matesprit}) to \cref{eq:GR_t}~\cite{Yu2026}.

Since ESPRIT returns whatever exponents reproduce the data, unphysical terms may appear and produce artifacts in $G^{R}(t)$ or in the spectral function. We therefore apply the filtering procedure of Ref.~\cite{Yu2026} to each $(m,\sigma)$ block. When an exponential $e^{s^{\gtrless}_{l}t}$ fitted on the complex-time contour is rotated back to the real-time axis, it becomes $e^{s^{\gtrless}_{l}e^{\pm i\theta}t}$; to prevent a divergence for $t\to\infty$,
\begin{equation}
  |e^{s^{\gtrless}_l e^{\pm i\theta} \Delta t}| \leq 1
  \label{eq:filter}
\end{equation}
must hold, and poles violating this condition are removed. Likewise, every pole $\xi_l$ of \cref{eq:pole_GR_t} must satisfy $\mathrm{Im}\,\xi_l\leq 0$, since $G^{R}(\omega)$ is analytic in the upper half plane, and poles violating this causality condition are removed as well. After the filtering, the weights of the remaining poles are determined anew from the overdetermined Vandermonde system, so that the weight of the discarded poles is redistributed among the remaining ones. As a safeguard, the positive semidefinite projection of \cref{sec:matrixesprit} can also be applied to the self-energy, to the local Green's function obtained from \cref{eq:selfcons}, to the mixed hybridization function, and to the weight matrices of the bath discretization (Appendix~\ref{app:disc}).

\section{Details of the matrix-valued ESPRIT}\label{app:matesprit}
In the matrix-valued ESPRIT~\cite{Zhang2024}, the $N_c\times N_c$ Green's function matrices sampled on an equidistant time grid of $N$ points are flattened into vectors of length $N_c^{2}$ and arranged into a Hankel matrix. A singular value decomposition of this matrix yields $e^{s_l\Delta t}$, and the vector-valued weights follow from an overdetermined least-squares Vandermonde system. The number of exponents $M$ is set by a tolerance $\epsilon$ on the relative singular values, which is chosen in the interval $[10^{-7},10^{-2}]$, and the auxiliary parameter of the Hankel matrix is set to $L=2N/5$. The individual steps are as follows.

An $N_c\times N_c$ matrix-valued function is approximated by a sum of exponentials with matrix-valued weights, as in \cref{eq:ansatz},
\begin{equation}
    \mathbf{f}(t)\approx\sum_{l=1}^{M}\mathbf{R}_{l}\,e^{s_{l}t},
    \label{eq:esprit_model}
\end{equation}
where $\mathbf{f}(t)$ is sampled on an equidistant time grid $t_{i}=i\Delta t$ with $i=0,1,\cdots,N-1$. Instead of working with the matrices directly, we flatten $\mathbf{f}(t_{i})$ and $\mathbf{R}_l$ into column vectors $\vec{f}(t_{i})$ and $\vec{R}_l$ of length $N_c^2$. With the auxiliary parameter $L$, we arrange the data into a Hankel matrix $\bm{H}$ of size $N_c^2(N-L)\times(L+1)$ and compute its singular value decomposition,
\begin{equation}
    \bm{H}=\begin{pmatrix}\vec{f}\left(t_{0}\right) & \vec{f}\left(t_{1}\right) & \cdots  & \vec{f}\left(t_{L}\right) \\ \vec{f}\left(t_{1}\right) & \vec{f}\left(t_{2}\right) & \cdots  & \vec{f}\left(t_{L+1}\right) \\ \vdots  & \vdots  & \ddots  & \vdots  \\ \vec{f}\left(t_{N-L-1}\right) & \vec{f}\left(t_{N-L}\right) & \cdots  & \vec{f}\left(t_{N-1}\right)\end{pmatrix}=\bm{U}\bm{S}\bm{W},
    \label{eq:hankel}
\end{equation}
where $\bm{U}$ and $\bm{W}$ are unitary matrices of size $N_c^2(N-L)\times N_c^2(N-L)$ and $(L+1)\times (L+1)$, respectively, and $\bm{S}$ is a rectangular diagonal matrix of size $N_c^2(N-L)\times (L+1)$ whose diagonal elements $\sigma_0\geq\sigma_1\geq\cdots\geq\sigma_L\geq0$ are the singular values of $\bm{H}$. 

The number of exponents $M$ is the smallest index for which $\sigma_{M}/\sigma_0<\epsilon$. From the first $M$ rows of $\bm{W}$, we define
\begin{equation}
    {\bm{W}}_{s}=\bm{W}\left(1:M,\; 1+s:L+s\right),\qquad s=0,1,
    \label{eq:Ws}
\end{equation}
and the exponents $\{s_l\}$ follow from the eigenvalues $\{x_l\}$ of the $M\times M$ matrix 
\begin{equation}
    \bm{F}=\big(\bm{W}_{0}^{T}\big)^{+}\bm{W}_{1}^{T},\qquad x_{l}=e^{s_{l}\Delta t},
    \label{eq:Fmat}
\end{equation}
where $+$ denotes the pseudoinverse and $T$ is the transpose.

Finally, the vectorized weights $\vec{R}_l$ are obtained from the overdetermined Vandermonde system
\begin{equation}
    \begin{pmatrix}\vec{f}{\left(t_{0}\right)}^{T} \\ \vec{f}{\left(t_{1}\right)}^{T} \\ \vdots  \\ \vec{f}{\left(t_{N-1}\right)}^{T}\end{pmatrix}=\begin{pmatrix}1 & 1 & \cdots  & 1 \\ x_{1} & x_{2} & \cdots  & x_{M} \\ \vdots  & \vdots  & \ddots  & \vdots  \\ x_{1}^{N-1} & x_{2}^{N-1} & \cdots  & x_{M}^{N-1}\end{pmatrix}\begin{pmatrix}{\vec{R}}_{1}^{T} \\ {\vec{R}}_{2}^{T} \\ \vdots  \\ {\vec{R}}_{M}^{T}\end{pmatrix},
    \label{eq:vandermonde}
\end{equation}
which is solved in the least-squares sense.

This procedure is applied separately to the complex-time greater and lesser Green's functions $G^{\gtrless}_{jj'}(t;\theta)$, which yields the representation of \cref{eq:ansatz}, and once more to the retarded Green's function of \cref{eq:GR_t}, which yields the poles and weights of \cref{eq:pole_rep}. After the filtering of Appendix~\ref{app:esprit}, only \cref{eq:vandermonde} is solved again with the remaining exponents.

\section{Three-orbital single-impurity Anderson model}\label{app:3orbsiam}
\begin{figure}[htbp]
    \centering
    \includegraphics[width=0.9\linewidth]{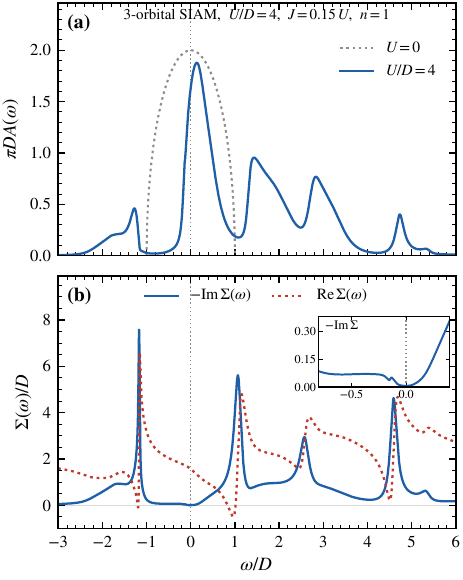}
    \caption{Spectral function and self-energy of the three-orbital single-impurity Anderson model with $U/D=4$, $J=0.15U$, and $n\simeq1$. (a) Spectral function $\pi DA(\omega)$ of a single $(m,\sigma)$ (solid line) and semicircular density of states of the non-interacting Bethe lattice (dotted line). (b) Imaginary part $-\mathrm{Im}\,\Sigma(\omega)$ (solid line) and real part $\mathrm{Re}\,\Sigma(\omega)$ (dotted line) of the self-energy; the inset shows $-\mathrm{Im}\,\Sigma(\omega)$ near $\omega=0$. The vertical dotted line marks the Fermi level $\omega=0$.}
    \label{fig:3orbsaim}
\end{figure}

We solve the three-orbital Anderson impurity model with $U/D=4$, $J=0.15U$, and half bandwidth $D=1$ at the filling $n\simeq1$, the model parameters of Fig.~8 of Ref.~\cite{Yu2026}. The bath is represented in the double-chain geometry with $N_b=399$ bath sites per $(m,\sigma)$, and the bond dimension is $\chi=150$. The complex-time evolution uses $\theta=0.2$ and $D\Delta t=0.1$ up to $Dt_{\max}=50$, the broadening $\eta=10^{-10}$, and the ESPRIT tolerance is $\epsilon=10^{-5}$.

\cref{fig:3orbsaim}(a) shows the spectral function $\pi DA(\omega)$. In contrast to the two-orbital model, several peaks appear: one near the Fermi level at $\omega/D\simeq 0.14$ and further peaks at $\omega/D\simeq-1.28$, $1.43$, $2.83$, and $4.73$. This peak structure agrees with Fig.~8 of Ref.~\cite{Yu2026}, where it was interpreted as the characteristic multiplet structure associated with the Kanamori interaction, with peaks corresponding to hole excitations, high-spin states, low-spin states, and Hubbard bands.

\cref{fig:3orbsaim}(b) shows the self-energy. The imaginary part $-\mathrm{Im}\,\Sigma(\omega)$ has its minimum at $\omega=0$, with $-\mathrm{Im}\,\Sigma(0)\simeq6.2\times10^{-3}D$ [inset of \cref{fig:3orbsaim}(b)], and exhibits distinct peaks at $\omega/D\simeq-1.17$, $1.07$, $2.57$, and $4.59$, which lie close to the minima between the peaks of the spectral function in \cref{fig:3orbsaim}(a).

\bibliography{ref}

\end{document}